\documentclass[
  aps,
  prapplied,
  reprint,
  superscriptaddress
]{revtex4-2}

\usepackage{hyperref}
\usepackage{graphicx}
\usepackage{siunitx}
\graphicspath{{./figures/}}

\usepackage{lineno}

\begin{document}

\title{Improving signal-to-noise ratios in pump-probe spectroscopy on light-sensitive samples by adapting pulse repetition rates}

\author{Matthias C. Velsink}
\email[Contact author: ]{m.velsink@arcnl.nl}
\author{Maksym Illienko}
\author{Komal Chaudhary}
\affiliation{Advanced Research Center for Nanolithography (ARCNL), Science Park 106, 1098 XG Amsterdam, The Netherlands}
\author{Stefan Witte}
\email[Contact author: ]{s.witte@arcnl.nl}
\altaffiliation[Present address: ]{Imaging Physics Department, Faculty of Applied Sciences, Delft University of Technology, Lorentzweg 1, 2628 CJ Delft, The Netherlands}
\affiliation{Advanced Research Center for Nanolithography (ARCNL), Science Park 106, 1098 XG Amsterdam, The Netherlands}
\affiliation{Department of Physics and Astronomy, Vrije Universiteit, De Boelelaan 1081, 1081 HV Amsterdam, The Netherlands}

\begin{abstract}
Ultrafast optical pump-probe spectroscopy is a powerful tool to study dynamics in solid materials on femto- and picosecond timescales.
In such experiments, a pump pulse induces dynamics inside a sample by impulsive light-matter interaction, resulting in dynamics that can be detected using a time-delayed probe pulse.
In addition to the desired dynamics, the initial interaction may also lead to unwanted effects that may result in irreversible changes and even damage.
Therefore, the achievable signal strength is often limited by the pumping conditions that a sample can sustain.
Here we investigate the optimization of ultrafast photoacoustics in various solid thin films.
We perform systematic experiments aimed at maximizing the achievable signal-to-noise (SNR) ratio in a given measurement time while limiting sample damage.
By varying pump and probe pulse energies, average pump fluence, and repetition rate, we identify different paths towards optimal SNR depending on material properties.
Our results provide a strategy for the design of pump-probe experiments, to optimize achievable SNR for samples in which different damage mechanisms may dominate.
\end{abstract}

\maketitle

\section{Introduction}
Pump-probe spectroscopy is a versatile technique for a wide range of applications, often aimed at material characterization~\cite{maiuri2020} or at the generation of a (material-specific) optical response in imaging studies~\cite{fischer2016}.
The concept of pump-probe spectroscopy is to induce dynamics in a sample through ultrafast light-matter interaction, which can subsequently be probed via a time-dependent optical response.
Specificity is a challenge in pump-probe experiments, as the optical pumping can induce spurious effects such as heating in addition to the intended dynamics, resulting in background signal, unwanted sample variation and even damage.
Finding optimal parameters for a pump-probe experiment is therefore typically challenging and strongly sample-dependent.
While nonlinear optical excitation schemes profit from higher pulse energy, various damage mechanisms based on multi-photon pathways also become increasingly important~\cite{vonderlinde2000,rethfeld2017}, providing a limit in applicable single pulse energy.
At the same time, thermal effects typically provide a limit on average power.
For pulsed illumination, cumulative build-up effects caused by multiple consecutive pulses can also play a role in sample damage, further complicating the choice of parameters.

The repetition rate is shown to be important in femtosecond micromachining~\cite{gattass2006}, providing a balance between single-shot ablation and thermally induced material changes.
Fluorescence lifetime imaging can be performed with synchronized lasers, by detecting stimulated emission induced by a probe laser with a variable time delay~\cite{dong1995,buehler2000}.
This concept can even be extended to imaging by detecting molecules via stimulated emission~\cite{min2009}, and for high-sensitivity stimulated Raman scattering microscopy~\cite{freudiger2008}, and coherent anti-Stokes Raman spectroscopy~\cite{ideguchi2013}.
In all these applications, sample damage is a limiting factor, and there are limitations on both the acceptable pulse energy and average power incident on a sample.
The influence of repetition rate on fluorescence microscopy was shown to be large for specific fluorophores, as electronic relaxation timescales play an important role~\cite{donnert2007}.
Two-photon excitation obviously benefits from a higher pulse energy at reduced repetition rate (maintaining constant average power)~\cite{boguslawski2024}.
Systematic studies on multiphoton brain imaging showed that heating effects can be significant and limiting~\cite{podgorski2016}.
Further work highlighted the complex role of repetition rate on photodamage in live-cell multiphoton microscopy, showing reduced photobleaching and damage at sub-MHz repetition rate compared to \qty{100}{MHz}~\cite{macias-romero2016}.

Photoacoustic imaging and spectroscopy are pump-probe techniques in which an optical pulse induces acoustic signals in a sample, which can subsequently be detected in a spatially resolved way.
The acoustic signal is generated by a rapid local temperature increase, initiated by a laser pulse that heats the electron gas in the sample~\cite{thomsen1986}.
In biomedical applications, this acoustic pulse is detected using an ultrasound transducer~\cite{wang2016a}, and can even be used to assess local temperature~\cite{wang2014}.
In thin-film diagnostics, an optical probe pulse detects strain-dependent reflectivity changes that encode local material properties~\cite{zhang2020,illienko2024} and can be used for imaging through opaque layers~\cite{antoncecchi2020}.
The signal generation in photoacoustics is accompanied by sample heating, as is the case for most condensed matter pump-probe experiments.
The heating mechanisms depend on the electron-lattice interactions~\cite{tas1994}, and can become highly complex for nanostructured samples~\cite{hoogeboom-pot2015,frazer2019,beardo2021}.

In this work, we explore the optimization of signal-to-noise ratio (SNR) in photoacoustic experiments on thin films of various materials.
We identify different optimum operating regimes that are determined by specific material properties.
Using our versatile pump-probe setup based on modulated asynchronous optical sampling (MASOPS)~\cite{velsink2023}, we achieve shot-noise-limited conditions for a wide range of repetition frequencies, pulse energies and average powers, allowing a systematic optimization of SNR for very different thin film samples.
We show that by choosing appropriate parameters, SNR gains of several orders of magnitude can be achieved within a set measurement time and/or for given damage thresholds.
Our approach provides a framework for optimizing pump-probe experiments in other applications beyond photoacoustics as well.

\subsection{Theoretical scaling of the signal-to-noise ratio in pump-probe spectroscopy}
In many ultrafast pump-probe experiments, the pump-induced change in the sample increases linearly with either the pump pulse energy or fluence.
In the case of photoacoustics, where a pump pulse induces strain in a material, the signal increases linearly with the pump fluence $F_\mathrm{pump}$ if the spot size of the probe is much smaller than that of the pump.
This scaling holds because the increase in lattice temperature of the sample is linear in pump fluence, and the generated thermal expansion is linear in lattice temperature.
For a weak strain pulse, the probe reflectivity change $\Delta R$ is also linear.
Combined, the signal per pulse ($s$) should scale as $s \propto \Delta R \cdot E_\mathrm{probe} \propto F_\mathrm{pump} \cdot E_\mathrm{probe}$, with $E_\mathrm{probe}$ the probe pulse energy.
For an exposure time $T$ and pump repetition rate $\Gamma$, the total integrated signal $S \propto F_\mathrm{pump} \cdot E_\mathrm{probe} \cdot \Gamma \cdot T$.

For a shot-noise-limited experiment, the noise per pulse ($n$) will scale as $n \propto
\sqrt{E_\mathrm{probe}}$.
Because the noise in each pulse is statistically independent from subsequent pulses, the total integrated noise $N \propto \sqrt{E_\mathrm{probe} \cdot
\Gamma \cdot T \cdot \gamma}$, with $\gamma$ the number of probe pulses per pump pulse.
Often, there is one probe pulse for every pump pulse, and $\gamma = 1$.
However, in experiments using lock-in detection typically $\gamma =
2$, as then probe pulses are detected both with and without pump.

Together, this yields an amplitude signal-to-noise ratio (SNR) of
\begin{equation}\label{eq:snr_scaling}
  \mathrm{SNR} = S/N
               = C \cdot F_\mathrm{pump} \cdot \sqrt{\frac{E_\mathrm{probe} \cdot \Gamma
                 \cdot T}{\gamma}},
\end{equation}
where $C$ is a parameter that depends on pump and probe wavelengths and the sample material.
Note that $C$ is also influenced by the measurement bandwidth $\Delta f$.
The bandwidth should be chosen such that the signal is not attenuated, whilst limiting noise, as noise increases proportional to $\sqrt{\Delta f}$.
If and how the bandwidth can actually be adjusted depends on the experiment.

In case the experiment is not shot-noise limited, the noise per pulse is a constant, and the number of probe pulses per pump pulse is irrelevant.
The total integrated noise is then $N \propto \sqrt{\Gamma \cdot T}$, and Eq.~\ref{eq:snr_scaling} simplifies to
\begin{equation}\label{eq:snr_scaling_not_shot}
  \mathrm{SNR} = C \cdot F_\mathrm{pump} \cdot E_\mathrm{probe}
                 \cdot \sqrt{\Gamma \cdot T}.
\end{equation}
Note that Eq.~\ref{eq:snr_scaling_not_shot} only holds if probe laser intensity noise also plays no role, and background noise sources dominate.





\section{Experimental setup}
\begin{figure}[b]
  \includegraphics{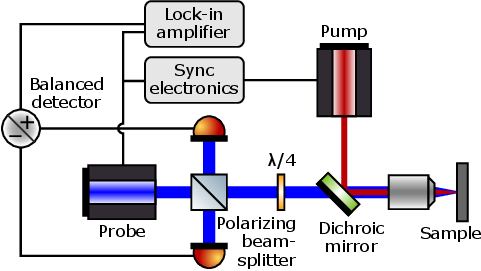}
  \caption{Schematic of the experimental setup.
  The pump and probe laser are electronically synchronized.
  A dichroic mirror combines the two beams, which are then focused onto the sample with a microscope objective.
  The reflected probe light is analyzed with a balanced detector and lock-in amplifier.
  Both pump and probe can be pulse-picked to lower repetition rates.}
  \label{fig:setup}
\end{figure}

To study the effect of pulse picking on achievable SNR, we generate and detect ultrafast photoacoustics in three different solid thin films.
The setup for this experiment is illustrated in Fig.~\ref{fig:setup}.
The measurement concept is based on Modulated Asynchronous Optical Sampling (MASOPS)~\cite{velsink2023}.
Two separate pump (Menlo Systems Orange) and probe (Menlo Systems C-Fiber 780) lasers are electronically synchronized with a variable pump-probe delay.
The pump and probe are focused on the same spot on the sample, after which the reflected probe light is analyzed with a balanced detector and lock-in amplifier (Zurich Instruments UHFLI).
More details are provided in~\cite{velsink2023}.
Nominally, the pump and probe run at \qty{50}{MHz} and \qty{100}{MHz}, respectively.
Note that in our measurement scheme, the repetition frequency of the probe is two times higher than the pump to enable lock-in detection at the pump repetition frequency.
However, both lasers can be pulse-picked to reduce their repetition rate.
For the probe, this is achieved with an external Pockels cell, and for the pump with an internal AOM with post-amplifier.
The pump can therefore have higher pulse energy when pulse-picked, but the probe cannot.

The pump pulses have a center wavelength of \qty{1030}{nm} and a pulse duration of \qty{180}{fs}, while the probe wavelength is \qty{780}{nm}.
The probe pulses have a nominal duration of \qty{80}{fs}, although the external pulse picker introduces some dispersion that is corrected using chirped mirrors.
The $1/e^2$ spot radius of the probe on the sample is approximately \qty{1.0}{\micro m}.
Because the pulse picked output of the pump laser has a different beam diameter, we place a \qty{1.5}{mm} pinhole in the pump path to have a consistent pump beam diameter.
The resulting picked and unpicked pump $1/e^2$ spot radii in focus are \qty{5.5}{\micro m} and \qty{5.2}{\micro m}, respectively.
For the unpicked probe, we use a fast balanced detector (\qty{500}{MHz}, \qty{5}{kV/A}, Femto HBPR), but for the picked probe we use a slower detector with lower noise (\qty{8}{MHz}, \qty{40}{kV/A}, Koheron PD10B).
We can typically achieve shot-noise limited detection with both detectors, but will correct for any excess (detector) noise in the analysis.

Normally, ASOPS becomes inefficient at lower repetition rates because of the increased time interval between pulses, but with the added modulation there is no loss of measurement efficiency~\cite{velsink2023}.
We therefore do not recommend reducing repetition rates, unless MASOPS or another form of electronically controlled optical sampling~\cite{kim2010} is available.

\subsection{Pulse picking considerations}\label{sec:limitations}
In our experiments, we reduce the pump repetition rate to \qty{1}{MHz} when pulse picking.
Ideally, the probe should then be picked at \qty{2}{MHz} to still use our lock-in amplifier scheme.
However, as the switching time of our Pockels cell is insufficient for picking single pulses, we instead pick pulse pairs from the probe laser, with a time delay of \qty{10}{ns} between these pulses.
Moreover, our Pockels cell driver employs a bipolar voltage switching scheme, which introduces an asymmetry in beam pointing between consecutive pulse pairs~\cite{palatchi2021}.
We therefore pick probe pulse pairs at \qty{4}{MHz}, so that the pump pulses always align with the same pulse pair "polarity".
Lock-in detection is then still performed at \qty{1}{MHz}.

Because of this combination of factors, we have 8 probe pulses for every pump pulse in the pulse-picked experiment ($\gamma = 8$ in Eq.~\ref{eq:snr_scaling}).
Shot noise therefore increases by a factor $\sqrt{4} = 2$ compared to the ideal \qty{2}{MHz} picking, but can be corrected for in the analysis to emulate a setup without such practical limitations.

\subsection{Sample selection}
We generate and detect laser-induced ultrafast photoacoustics in three different samples.
Two are freestanding thin membranes (Luxel), one aluminum (\qty{394}{nm} thick), and the other tantalum (\qty{198}{nm} thick).
The third sample is a piece of silicon wafer covered by a first layer of \qty{50}{nm} SiO2 and a top layer of \qty{600}{nm} opaque amorphous carbon.
We chose these three samples because they have very different laser-induced damage mechanisms and thresholds.
For pulse-energy-limited samples, we expect pulse picking to not improve or even decrease SNR.
However, for average power limited samples, increasing pulse energy by pulse picking should increase SNR.

\section{Results and discussion}
For all three samples under consideration, we select a range of different measurement parameters in such a way that specific properties can be compared directly.
For example, measurements with different repetition frequencies at identical average power allow a comparison of the influence of single-pulse energy on SNR under constant thermal load.
A table is given for each sample with an overview of measurements with their respective parameter sets.

\subsection{Aluminum}
Aluminum thin films have a high damage threshold and good strain-optic sensitivity at our probe wavelength, allowing measurements with a good SNR for a wide range of experimental parameters.
We therefore use it to show the scaling of both signal and noise as a function of pump fluence, probe energy, and repetition rate.
Table~\ref{tab:alu} contains an overview of the different parameter combinations used in the experiments.
\begin{table}[b]
  \caption{Aluminum measurement parameters}
  \begin{ruledtabular}
    \begin{tabular}{c c c c c}
      &
        \shortstack{Pump rate \\ \unit{(MHz)}} &
        \shortstack{Pump fluence \\ \unit{(J/m^2)}} &
        \shortstack{Probe rate \\ \unit{(MHz)}} &
        \shortstack{Probe  energy \\ \unit{(pJ)}} \\
      \hline
      1 & 50 & 4.6 & 100 & 4.4 \\
      2 & 50 & 4.6 & 100 & 13 \\
      3 & 1  & 4.1 & 100 & 13 \\
      4 & 1  & 205 & 8   & 13 \\
      5 & 1  & 205 & 8   & 55 \\
    \end{tabular}
  \end{ruledtabular}
  \label{tab:alu}
\end{table}
We choose these parameters such that we have measurements with approximately equal pump/probe powers, as well as pump/probe energies.

\begin{figure}
  \includegraphics{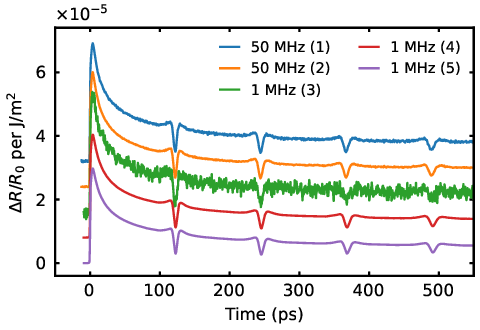}
  \caption{Transient reflectivity curves for a \qty{394}{nm} thick aluminum thin film at \qty{32}{s} integration time, normalized by pump fluence.
  The laser parameters for each measurement are given in Table~\ref{tab:alu}.
  Each trace is vertically offset by \num{0.8e-5} for clarity.
  The picked measurement (5), with equal average powers for pump and probe as the unpicked measurement (1), has the highest SNR.}
  \label{fig:alu_curves}
\end{figure}

The transient pump-probe reflectivity curves measured for these different sets of parameters are shown in Fig.~\ref{fig:alu_curves}.
For all measurements, the total integration time was \qty{32}{s}.
These curves show the typical time-dependent response of aluminum films~\cite{zhang2020,velsink2023}.
Heat transport into the sample causes a fast exponential decay, while the back-and-forth travelling of the acoustic pulse shows up as periodic dips in reflectivity~\cite{matsuda2015}.
As expected, the signal strength is the same when normalized by the pump fluence.

\subsubsection{SNR analysis}\label{sec:alu_SNR}
We analyze the noise in the measurement by Welch's method~\cite{welch1967}, using \qty{50}{ps} long Hann-windowed segments with \qty{50}{\percent} overlap.
This procedure is similar to the approach we developed in earlier work~\cite{velsink2023}.
We then correct for the filter response of the lock-in amplifier, average the white noise density above \qty{400}{GHz}, and weigh it by \qty{200}{GHz} to estimate the total root-mean-square (RMS) $\Delta R/R_0$ noise within the measurement.
In order to not be affected by experimental background noise, we also measure the noise density without pump and probe and correct for this excess noise.
Still, our experiment is almost fully shot-noise limited, as the excess noise amplitude is only \qty{14}{\percent} of the total noise for the unpicked measurement (1) with the lowest probe power.

\begin{figure}
  \includegraphics{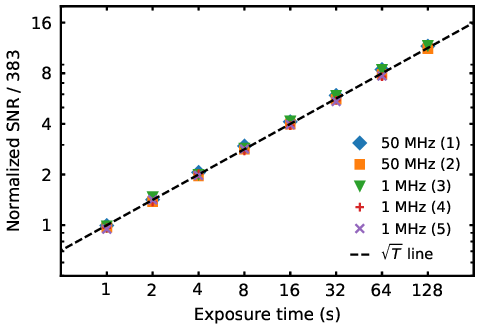}
  \caption{The SNR for the first aluminum echo (see Fig.~\ref{fig:alu_curves}) over a \qty{200}{GHz} bandwidth for different exposure times.
  The values are normalized by the expected scaling factor and divided by $C \approx 383$.
  As expected, the SNR scales proportional to $\sqrt{T}$.}
  \label{fig:alu_snr_norm}
\end{figure}

The signal in our measurements is defined as the peak-to-peak amplitude of the first acoustic echo at a pump-probe delay time of \qty{120}{ps}.
To get the most accurate determination of the signal amplitude, we use the echo amplitude of the longest exposure time to define the signal strength in the SNR analysis for all exposure times.
From Eq.~\ref{eq:snr_scaling}, we expect that normalizing the SNR of each measurement by $F_\mathrm{pump} \cdot \sqrt{E_\mathrm{probe} \cdot \Gamma \cdot T / \gamma}$ should result in overlapping curves scaling as $C \cdot \sqrt{T}$.
As shown in Fig.~\ref{fig:alu_snr_norm}, this is indeed the case, with $C \approx 383$.
We therefore conclude that Eq.~\ref{eq:snr_scaling} describes the SNR scaling in our experiment well.
In general, for any shot-noise limited pump-probe experiment in which the signal scales linearly with pump fluence and probe pulse energy, Eq.~\ref{eq:snr_scaling} should hold.

\subsection{Tantalum}
Compared to the aluminum sample, the tantalum membrane damages very easily.
From systematic tests covering a range of different laser parameters, we find that the damage does not significantly depend on pulse energy, but mainly on average power.
This observation indicates that thermal effects (melting) are the main damage mechanism.
At an average pump intensity of \qty{46}{MW/m^2}, we find that damage remains insignificant for measurement times beyond one minute.
This average power limit is similar for both the unpicked and picked pump, meaning the pump pulse energy can be around 50 times higher when pulse-picking at \qty{1}{MHz} repetition rate.
This energy increase should boost SNR proportionally if the probe power is also kept constant.

\begin{figure}
  \includegraphics{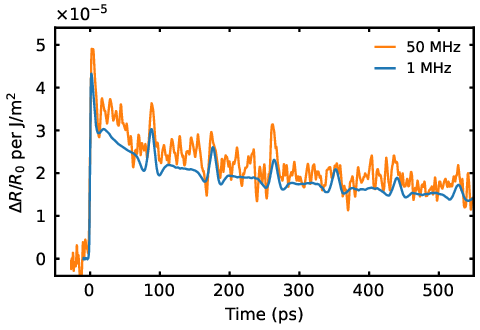}
  \caption{Transient reflectivity curves for a \qty{198}{nm} thick tantalum thin film at \qty{8}{s} integration time, normalized by pump fluence.
  The average pump and probe power are equal in both measurements.
  Pulse picking clearly increases SNR.}
  \label{fig:tanta_curves}
\end{figure}

The comparison between the unpicked and picked pump-probe experiments at roughly equal pump intensity (\qty{46}{MW/m^2}) and probe power (\qty{220}{\micro W}) is plotted in Fig.~\ref{fig:tanta_curves}.
The exact parameters for both measurements are given in Table~\ref{tab:tanta}.
\begin{table}[b]
  \caption{Tantalum measurement parameters}
  \begin{ruledtabular}
    \begin{tabular}{c c c c}
        \shortstack{Pump rate \\ \unit{(MHz)}} &
        \shortstack{Pump fluence \\ \unit{(J/m^2)}} &
        \shortstack{Probe rate \\ \unit{(MHz)}} &
        \shortstack{Probe energy \\ \unit{(pJ)}} \\
      \hline
      50 & 0.92 & 100 & 2.2 \\
      1  & 41   & 8   & 28 \\
    \end{tabular}
  \end{ruledtabular}
  \label{tab:tanta}
\end{table}
The pump intensity of the pulse picked measurement is slightly lower (\qty{41}{MW/m^2}) due to beam size differences.

The time-dependent reflectivity changes induced in the tantalum thin film show a similar thermal decay as the aluminum samples, but the acoustic echoes have the opposite sign, i.e. a transient reflectivity increase.
It is clearly visible that the pulse picked measurement has much better SNR.
This "free" gain in SNR is perhaps counterintuitive, given that the average pump and probe powers are equal in both measurements.
However, the signal is effectively proportional to a product of pump fluence and probe energy, which scales faster than the linear loss of repetition rate.

\begin{figure}[b]
  \includegraphics{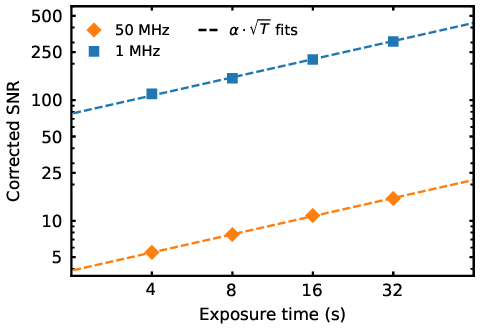}
  \caption{The SNR for the first tantalum echo (see Fig.~\ref{fig:tanta_curves}) over a \qty{200}{GHz} bandwidth for different exposure times.
  The $\sqrt{T}$ scaling constant for the \qty{50}{MHz} measurements is $\alpha = 2.7$, and $\alpha = 55$ for \qty{1}{MHz}.
  Pulse picking thus improves SNR by a factor of 20.}
  \label{fig:tanta_snr}
\end{figure}

For a quantitative comparison, the SNR without normalization is analyzed in the same way as in section~\ref{sec:alu_SNR}, and plotted in Fig.~\ref{fig:tanta_snr}.
The SNR is corrected for excess noise and the pulse-picked SNR is multiplied by 2 (see section~\ref{sec:limitations}).
Both picked and unpicked measurements show the $\sqrt{T}$ scaling as predicted by Eq.~\ref{eq:snr_scaling}.
Pulse picking is found to improve SNR by a factor 20 in this experiment.
The picked pump fluence is about 45 times higher, and ideally the SNR increase would reach a similar value if the probe repetition rate would be similarly reduced while keeping probe power equal.
However, because of the probe pulse picking limitations described in section~\ref{sec:limitations}, the probe pulse energy at equal power is a factor 4 lower than it could be for ideal pulse picking at \qty{2}{MHz}.
That halves the achievable SNR improvement.
Furthermore, small alignment differences between the pump and probe can also influence the signal strength.
Therefore, the observed SNR increase is within expectations, and the scaling from Eq.~\ref{eq:snr_scaling} holds.
We expect the SNR improvement to increase further for even lower repetition rates, until the damage is dominated by single pulses and not by average power alone.

\subsection{Amorphous carbon}
Amorphous (hydrogenated) carbon is a material that is used in semiconductor manufacturing as an etch hard mask for high-aspect ratio features.
Ultrafast photoacoustics can be used to characterize these masks~\cite{dai2020}.
In such thin-film metrology applications, maximizing SNR is important for rapid acquisition, whilst keeping laser-induced damage to a minimum.
Unlike tantalum, which damages by sudden melting, amorphous carbon is modified (and eventually damages) by thermal annealing due to the laser-induced heat.
This annealing process increases opacity and reduces sound velocity~\cite{sattel1997,arlein2008}.

\begin{figure}
  \includegraphics{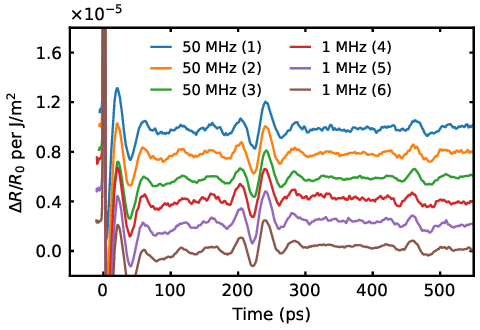}
  \caption{Transient reflectivity curves for a \qty{600}{nm} thick amorphous carbon film recorded with \qty{1}{s} integration time for (1) and (4), \qty{4}{s} for (2) and (5), and \qty{16}{s} for (3) and (6), normalized by pump fluence.
  The details for each measurement are given in Table~\ref{tab:aC}.
  Each trace is vertically offset by \num{0.2e-5} for clarity.}
  \label{fig:aC_curves}
\end{figure}

To characterize the achievable SNR at different levels of damage, we performed ultrafast photoacoustics measurements in the amorphous carbon layer at different laser intensities and total exposure times.
The different laser parameter settings are shown in Table~\ref{tab:aC}.
\begin{table}[b]
  \caption{Amorphous carbon measurement parameters}
  \begin{ruledtabular}
    \begin{tabular}{c c c c c}
        &
        \shortstack{Pump rate \\ \unit{(MHz)}} &
        \shortstack{Pump fluence \\ \unit{(J/m^2)}} &
        \shortstack{Probe rate \\ \unit{(MHz)}} &
        \shortstack{Probe energy \\ \unit{(pJ)}} \\
      \hline
      1 & 50 & 24  & 100 & 18 \\
      2 & 50 & 18  & 100 & 13 \\
      3 & 50 & 15  & 100 & 11 \\
      4 & 1  & 136 & 8   & 44 \\
      5 & 1  & 109 & 8   & 33 \\
      6 & 1  & 88  & 8   & 28 \\
    \end{tabular}
  \end{ruledtabular}
  \label{tab:aC}
\end{table}
A selection of the measured datasets are plotted in Fig.~\ref{fig:aC_curves}.
For these carbon films, the first echo arrives after around \qty{230}{ps}, but some oscillations are already visible before that.
These are Brillouin oscillations caused by optical interference between the reflected probe light from the front of the sample, and that reflected by the refractive index change from the echo inside~\cite{dai2020}.

\begin{figure}[b]
  \includegraphics{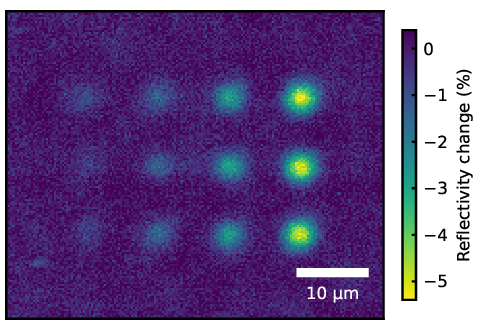}
  \caption{Change in reflectivity of the amorphous carbon film as measured using wide-field microscopy after pump exposure at increasing exposure time, for the pulse-picked case with parameters corresponding to measurement (6) in Table~\ref{tab:aC}.
  The observed reflectivity drop is interpreted as pump-induced damage.
  For the spots from left to right, exposure times are \qty{16}{s}, \qty{32}{s}, \qty{64}{s}, and \qty{128}{s}.
  Every exposure time is vertically repeated three-fold.
  The maximum damage is around \qty{5}{\percent}.}
  \label{fig:aC_dmg}
\end{figure}

To characterize the pump-induced modification or damage, we image the sample surface with a bright-field reflection microscope as shown in Fig.~\ref{fig:aC_dmg}, and assess the local relative reduction in reflectivity.
From these images, we retrieve the mean relative reflectivity drop in a \qty{6}{\micro m} diameter circle around the center of the spot, and compare datasets with similar amounts of sample modification.

\begin{figure}
  \includegraphics{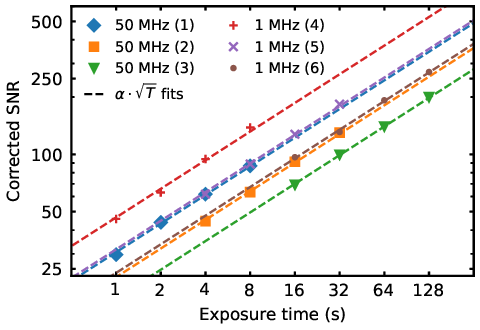}
  \caption{The SNR for the first amorphous carbon echo (see Fig.~\ref{fig:aC_curves}) over a \qty{200}{GHz} bandwidth for different exposure times and the 6 measurement parameter sets from Table~\ref{tab:aC}.
  In general, pulse picking gives better SNR at equal exposure times for these pump/probe powers.}
  \label{fig:aC_snr}
\end{figure}

The measured SNR, calculated via the method outlined in section~\ref{sec:alu_SNR} and corrected for excess noise and non-ideal pulse picking, is plotted in Fig.~\ref{fig:aC_snr}.
The scaling again follows Eq.~\ref{eq:snr_scaling}.
Pulse picking gives better SNR at equal exposure times when compared to not pulse picking.
Because damage increases with exposure time, the measurements with higher laser powers were limited in exposure time.
Similarly, the minimum exposure time is increased at lower laser powers until damage-induced reflectivity changes become detectable.

\subsubsection{SNR at the cost of damage}\label{sec:aC_SNR}
\begin{figure}[b]
  \includegraphics{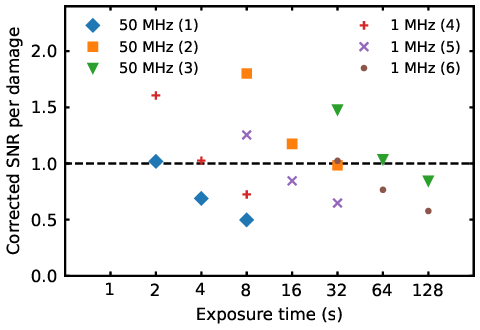}
  \caption{The same SNR as in Fig.~\ref{fig:aC_snr}, but now normalized by damage, for the three highest exposure times per group.
  All points are divided by the overall mean of \num{2.0e5} to compare the overall SNR-to-damage efficiency.
  At short exposure times, pulse picking is more efficient, but the higher repetition rate is better for longer exposure times.
  In general, slightly reducing fluence at similar exposure times can greatly decrease damage and therefore increase efficiency.}
  \label{fig:aC_snr_dmg}
\end{figure}

With the damage as defined above, we determine the SNR-to-damage efficiency as the SNR divided by the observed post-measurement reflectivity drop.
Because the shortest exposure times in each set of measurements result in little damage, we ignore them, as the reflectivity drop is close to 0 and sensitive to image noise.
To compare between sets of experimental settings, we normalize all the efficiency values by the global mean of all values, as plotted in Fig.~\ref{fig:aC_snr_dmg}.
So, the mean of all plotted normalized values is 1.
A value below 1 thus means worse efficiency than the average, and vice versa.
For each of the unpicked and pulse-picked experimental parameters, longer exposure times lead to worse efficiency.
This observation implies that damage scales faster than the $\sqrt{T}$ scaling of the SNR, which the measurements in appendix~\ref{app:aC_dmg_scaling} confirm.
However, the observed damage also increases rapidly with increased average power, faster than the SNR increases.
Therefore, measurements performed at lower average power and similar exposure time lead to a better SNR-to-damage efficiency.
These observations hold for both the unpicked and pulse-picked experiments.

However, at shorter exposure times, pulse picking has better efficiency, whereas at longer exposure times, not pulse picking is better.
In further tests, we have confirmed that damage scales faster with exposure time for the pulse-picked measurements than for the unpicked ones (see appendix~\ref{app:aC_dmg_scaling}).
Because of this difference in damage scaling with exposure time, there is a crossover between the measurement approaches.
For this specific sample, we find that at or below an exposure time of \qty{8}{s} pulse picking is more efficient.
Moreover, for ideal pulse picking with fewer redundant probe pulses, the pulse picked probe energy can be increased at the same average power, which should enhance SNR further.
Therefore, for amorphous carbon films, pulse picking to repetition rates of \qty{1}{MHz} enables faster ultrafast photoacoustics measurements.

\subsection{Discussion}
From these photoacoustics experiments on samples with so widely different properties, several general observations can be made.
The measurements confirm the expectation from the theory analysis, that increasing pump fluence and probe pulse energy lead to an SNR improvement that follows Eq.~\ref{eq:snr_scaling} under shot-noise-limited conditions.
However, increasing pump fluence can be achieved through increasing pulse energy or repetition frequency, and while both methods have an equal effect on SNR in principle, the dominant damage mechanism for a given sample determines which route is preferred.
For metallic samples such as Al and Ta, a viable optimization strategy is to first determine the single-pulse energy damage threshold.
Given the pump pulse energy limit, the maximum repetition rate is then determined by thermal damage limitations.

If the available pulse energy far exceeds the single-pulse damage threshold, fluence can of course also be limited by illuminating a larger area if the spatial properties of a sample are homogeneous.
For samples that are sensitive to thermal effects, this SNR optimization strategy typically favors repetition frequencies in the \qty{1}{MHz} range and below.
In practice, many detection systems also have a lower limit to the optimum repetition frequency range, below which slow drifts and ambient noise become significant compared to the shot noise limit.

In cases where damage (or more generally light-induced sample modification) does not scale linearly with fluence, defining the parameters that optimize SNR becomes even more challenging.
Our experiments on amorphous carbon showcase such a situation, for which a general scaling law to optimize SNR at a given damage limit cannot be defined.
Figure~\ref{fig:aC_snr_dmg} for instance shows that the optimum repetition rate changes for different total exposure time, and that for this specific material a higher SNR per damage can be achieved only for longer measurement times.
If an application sets specific boundary conditions for a measurement, such as a required measurement speed, an overview of the parameter space such as given in Fig.~\ref{fig:aC_snr_dmg} is highly beneficial.
While such parameters will be sample-dependent for nonlinear damage mechanisms, the analysis procedure introduced here gives a framework on how to determine the experimental conditions that optimize SNR in pump-probe spectroscopy.







\section{Conclusion}
We have investigated strategies to optimize the SNR in pump-probe spectroscopy experiments in the presence of damage mechanisms of varying origin.
Our results show the importance of adjusting especially the pulse repetition frequency to maximize SNR in a given measurement time, while limiting damage or other light-induced sample modifications.
Especially for samples with high sensitivity to average fluence, a proper choice of experimental parameters enables several orders of magnitude SNR advantage.
While the present work has focused on photoacoustic spectroscopy in thin solid films, the concepts and conclusions hold for a much wider range of time-resolved pump-probe spectroscopy methods and applications.
For instance, stimulated Raman spectroscopy and transient absorption spectroscopy have SNR equations very similar to Eqs.~\ref{eq:snr_scaling} and \ref{eq:snr_scaling_not_shot}, and therefore similar experimental design considerations apply.
As solution-based samples and biological specimen are often highly sensitive to thermal effects, we expect our conclusions to be applicable to such systems as well.

\begin{acknowledgments}
We acknowledge the support from the European Research Council (ERC-CoG 864016, project 3D-VIEW), and the Dutch Research Council NWO (TTW-HTSM 17960, project Orpheus).
This work was conducted at the Advanced Research Center for Nanolithography, a public-private partnership between the University of Amsterdam (UvA), Vrije Universiteit Amsterdam (VU), Rijksuniversiteit Groningen (RUG), the Dutch Research Council (NWO), and the semiconductor equipment manufacturer ASML.
\end{acknowledgments}

\appendix
\section{Amorphous carbon damage scaling}\label{app:aC_dmg_scaling}
\begin{figure}
  \includegraphics{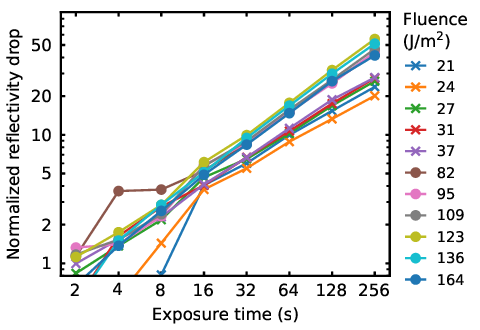}
  \caption{The normalized laser-induced damage for different fluences and at different exposure times.
  The curves with crosses are measured at the unpicked \qty{50}{MHz} repetition rate, and the curves with filled circles are measured at \qty{1}{MHz}.
  With a pulse-picked pump, the damage increases faster with exposure time than for the unpicked pump.}
  \label{fig:aC_damage_all}
\end{figure}

In order to understand the effect of repetition rate on the SNR-to-damage efficiency from section~\ref{sec:aC_SNR} on amorphous carbon better, we measure laser-induced damage at different fluences at both \qty{1}{MHz} and \qty{50}{MHz} pump rates.
Again, we quantify the damage by the average reflectivity drop in a circle around the pump spot, now with a \qty{8}{\micro m} diameter.
Here, we are interested in how the damage scales with exposure time, but not in the absolute damage.
Therefore, we fit an $\alpha \cdot T^\beta$ curve to the 4 longest exposure times ($T$) for each fluence, with $\alpha$ and $\beta$ the fit parameters.
We then divide the data for each fluence by $\alpha \cdot 2^\beta$ to normalize the trends to 1 at \qty{2}{s} integration time.
The resulting curves are shown in Fig.~\ref{fig:aC_damage_all}.
As before, at short exposure times for the lowest fluences, the damage is hard to quantify due to image noise.
Note that these measurements are without probe, so the overall damage is lower than for similar pump fluences as in section~\ref{sec:aC_SNR}.

It is clear that the damage increases faster with exposure time when pulse picking compared to not pulse picking.
The average $\beta$ for the unpicked measurements is 0.67, whereas for the picked measurements it is 0.79.
Both scale faster than $\sqrt{T}$ ($\beta = 0.5$), meaning the damage increases faster than SNR, as is visible in Fig.~\ref{fig:aC_snr_dmg}.
The $\alpha$ values have a similar range for both the picked (\qtyrange{0.04}{0.28}{\percent}) and unpicked (\qtyrange{0.07}{0.29}{\percent}) measurements.
Therefore, at shorter exposure times pulse picking can be better, as the damage also decreases faster than for the unpicked measurements when decreasing exposure times.

\bibliography{picking}

\end{document}